\documentclass[twocolumn,english,aps,prc,groupedaddress,amsmath,amssymb,showpacs]{revtex4}
\usepackage[T1]{fontenc}
\usepackage[latin9]{inputenc}
\usepackage{bm}
\usepackage{amsmath}

\makeatletter

\providecommand{\tabularnewline}{\\}

\@ifundefined{textcolor}{}
{%
 \definecolor{BLACK}{gray}{0}
 \definecolor{WHITE}{gray}{1}
 \definecolor{RED}{rgb}{1,0,0}
 \definecolor{GREEN}{rgb}{0,1,0}
 \definecolor{BLUE}{rgb}{0,0,1}
 \definecolor{CYAN}{cmyk}{1,0,0,0}
 \definecolor{MAGENTA}{cmyk}{0,1,0,0}
 \definecolor{YELLOW}{cmyk}{0,0,1,0}
 }

\makeatother

\usepackage{babel}
\begin{document}

\title{Symmetries in the $\bm{g_{9/2}}$ shell}

\author{L.Zamick}

\author{A. Escuderos}

\affiliation{Department of Physics and Astronomy, Rutgers University, Piscataway,
NJ 08854, USA}
\begin{abstract}
We consider symmetries which arise when two-body interaction matrix
elements with isopin $T=0$ are set equal to a constant in a single-$j$-shell
calculation. The nucleus $^{96}$Cd is used as an example. 
\end{abstract}
\maketitle

\section{Introduction}

The recent discovery of a $J=16^{+}$ isomer in $^{96}$Cd by Nara
Singh \textit{et al}.~\cite{setal11} was noted in a work on symmetries
in the $g_{9/2}$ shell by Zamick and Robinson~\cite{zr11}. In a
previous work by these authors~\cite{rz01}, the main emphasis was
on the $f_{7/2}$ shell, although the equations were written in a
general way so as to apply to any shell. In this work we elaborate
on the work in Ref.~\cite{zr11} by giving detailed wave functions
and energies.

First, however, we would like to point out that the $g_{9/2}$ shell
has, in recent years, been a beehive of activity both experimentally
and theoretically. For example, in contrast to the $f_{7/2}$ shell,
where the spectrum of three identical particles would be identical
to that of five particles (or three holes) in the single-$j$-shell
approximation, one can have different spectra in the $g_{9/2}$ shell.
It was noted by Escuderos and Zamick~\cite{ez06} that, with a seniority-conserving
interaction, the ($J=21/2^{+}$)--($J=3/2^{+}$) splitting (maximum
and mininum spins for three identical particles) was the same for
five identical particles as for three. However, with a $Q\cdot Q$
interaction (which does not conserve seniority), the splittings were
equal in magnitude but opposite in sign.

Even more interesting, it was noted in Ref.~\cite{ez06} that for
four identical particles (or holes, e.g. $^{96}$Pd) there are three
$J=4^{+}$ states---two with seniority 4 and one with seniority 2.
In general, seniority is not conserved in the $g_{9/2}$ shell. Despite
this fact, it was found that no matter what interaction was used,
one eigenstate emerged that was always the same for all interactions---this
was a seniority-4 state. The other two states were a mix of seniorities
2 and 4. A unique state also emerged for $J=6^{+}$. This observation
led to considerable activity with an intent to explain this behaviour
as seen in Refs.~\cite{z07,ezb05,ih08,zi08,q11,qxl12,t10}.

Another recently emerging topic has to do with $T=0$ pairing in which
the pairs are coupled to the maximum angular momentum, which for the
$g_{9/2}$ shell is $J=9^{+}$(see Refs.~\cite{cetal11,zi11,qbbcjlw11,xqblw12}).
In the Nature article~\cite{cetal11}, an experiment is presented
in which an almost equal-spaced spectrum is found in $N=Z$ $^{92}$Pd
for $J=0^{+},2^{+},4^{+}$, and $6^{+}$. This is well reproduced
by the shell model but also by $T=0$ pairing with maximum alignment
of pairs. The $2^{+}$ state is lower in $^{92}$Pd than in $^{96}$Pd,
which is not surprising because the latter is semi-magic. An interesting
observation in Ref.~\cite{zi11} is that, although the spectrum looks
vibrational, the $B(E2)$s might obey a rotational formula rather
than vibrational. It would be of interest to have experimental verification.

Recently the current authors addressed a different problem~\cite{ze12},
but one that has some implications for the above topic. We studied
the question of isomerism for systems of three protons and one neutron,
e.g. $^{96}$Ag. We found that for the upper half of a $j$ shell,
be it $f_{7/2}$ or $g_{9/2}$ shell, the $J=J_{\text{max}}$ two-body
matrix element is much more attractive than in the lower half. The
states in question have $J=7^{+}$ and $9^{+}$ respectively. This
was necessary to explain the isomeric behaviours of states in various
nuclei, e.g. why the lifetime of the $J=12^{+}$ state in $^{52}$Fe
is much longer than in $^{44}$Ti, or why the $J=16^{+}$ states in
$^{96}$Cd and the $J=15^{+}$ states in $^{96}$Ag are isomeric.
In that work, we also used the previously determined two-body matrix
elements of Coraggio \textit{et al.}~\cite{ccgi12} as well as our
own. With regards to the works of Refs.~\cite{cetal11,zi11,qbbcjlw11,xqblw12},
this suggests the $J=J_{\text{max}}$ pairing is a better approximation
in the upper half of a $j$ shell rather than the lower half.

It should be added that one of us had previously studied the problem
of the effects of varying two-body matrix elements in the $f_{7/2}$
region~\cite{zr02} and that one can learn a lot by studying the
explicit $f_{7/2}$ wave functions in Ref.~\cite{ezb05}, which are
based on previous works in Refs.~\cite{bmz63,gf63}.

But let us not lose sight of what this paper is about---partial dynamical
symmetries (PDS) in $^{96}$Cd. One obvious distinction of our work
relative to that of Refs.~\cite{cetal11,zi11,qbbcjlw11,xqblw12}
is that we are dealing with higher spins than the ones that they consider.
In the Nature article~\cite{cetal11}, they measure $J=0,2,4$, and
6 and also discuss $J=8$, whereas the PDS that we consider occur
for $J=11^{+}$ and beyond. Thus, our work may be regarded as different
but complimentary to theirs.

At the time of this writing, the $16^{+}$ isomer is the only known
excited state of $^{96}$Cd~\cite{setal11}. It decays to a $15^{+}$
isomer in $^{96}$Ag. In previous work by Escuderos and Zamick~\cite{ez06},
it was noted that the single-$j$ approximation for a few holes relative
to $Z=50$, $N=50$ works fairly well. However, they cautioned that
this approximation does not work at all for a few particles relative
to $Z=40$, $N=40$. The relevant shell is, of course, $g_{9/2}$.

\section{Theoretical framework}

The symmetry in question mentioned above comes from setting two-body
interaction matrix elements with isospin $T=0$ to be constant, whilst
keeping the $T=1$ matrix elements unchanged~\cite{zr11}. It does
not matter what the $T=0$ constant is as far as symmetries are concerned,
but it does affect the relative energies of states of different isospins.
Briefly stated, for the four-hole system $^{96}$Cd, we then have
a PDS, one which involves $T=0$ states with angular momenta that
do not exist for $T=2$ states in a pure $g_{9/2}$ configuration.
That is to say, the PDS will not occur for states with angular momenta
which can occur in the $(g_{9/2})^{4}$ configuration of four identical
particles ($^{96}$Pd). Although many 6-$j$ and 9-$j$ relations
were used in Refs.~\cite{zr11,rz01} to describe why these symmetries
are partial, one simple argument is illuminating. If a given angular
momentum can occur for, say, four $g_{9/2}$ neutron holes ($T=2$),
then there is a constraint on the $T=0$ states with the same angular
momentum---namely their wave functions have to be orthogonal to the
$T=2$ states. This constraint prevents the occurence of a PDS.

For qualifying $T=0$ states, the PDS consists of $J_{p}$ and $J_{n}$
being good dual quantum numbers. That is to say, the wave function
of a state will have only one ($J_{p},J_{n}$) and ($J_{n},J_{p}$).
Another way of saying this is that $J_{p}\cdot J_{n}$ is a good quantum
number. Furthermore, states with different total angular momentun
$J$ but with the same ($J_{p},J_{n}$) will be degenerate.

We have given a physical argument for the PDS. We can also explain
it mathematically. There are both off-diagonal and diagonal conditions.
The former is needed to explain why $(J_{p},J_{n})$ are good dual
quantum numbers. The reason is the vanishing of the $9j$-symbol 
\begin{equation}
\begin{Bmatrix}j & j & (2j-1)\\
j & j & (2j-1)\\
(2j-1) & (2j-3) & (4j-4)
\end{Bmatrix}=0\label{eq:9j}
\end{equation}
 Next we need diagonal conditions to explain why states with the same
$(J_{p},J_{n})$ are degenerate. These are given by 
\begin{equation}
\begin{Bmatrix}j & j & (2j-3)\\
j & j & (2j-1)\\
(2j-3) & (2j-1) & I
\end{Bmatrix}=\frac{1}{4(4j-5)(4j-1)}\label{eq:diag1}
\end{equation}
 for $I=(4j-4)$, $(4j-5)$, $(4j-7)$ and 
\begin{equation}
\begin{Bmatrix}j & j & (2j-1)\\
j & j & (2j-1)\\
(2j-1) & (2j-1) & I
\end{Bmatrix}=\frac{1}{2(4j-1)^{2}}\label{eq:diag2}
\end{equation}
 for $I=(4j-4)$, $(4j-2)$.

\section{Results}

How the partial dynamical symmetry manifests itself is best illustrated
by examining Tables~\ref{tab:96cd-intd} and \ref{tab:96cd-intd0}.
Here we use the two-body INTd matrix elements from Zamick and Escuderos~\cite{ze12}
to perform single-$j$-shell calculations of the energies and wave
functions of $^{96}$Cd. Actually, it does not matter what charge-independent
interaction is used to illustrate the symmetry that will emerge.

\begin{table*}[htb]
\caption{\label{tab:96cd-intd} Wave functions and energies (in MeV, at the
top) of selected states of $^{96}$Cd calculated with the INTd interaction
(see text).}

\begin{ruledtabular} %
\begin{tabular}{cccccccc}
\multicolumn{2}{l}{$J=11$} &  &  &  &  &  & \tabularnewline
 &  & 5.5640  & 5.6482  & 6.4693  & 6.6384  & 6.9319  & 8.1822 \tabularnewline
$J_{p}$  & $J_{n}$  & T=1 &  & \multicolumn{1}{r}{T=1} & \multicolumn{1}{r}{} & \multicolumn{1}{r}{$T=1$} & \multicolumn{1}{r}{$T=1$}\tabularnewline
4  & 8  & 0.4709  & -0.6359  & -0.2463  & 0.3092  & -0.4544  & 0.1051 \tabularnewline
6  & 6  & 0.2229  & 0.0000  & 0.8712  & 0.0000  & -0.3121  & -0.3065 \tabularnewline
6  & 8  & 0.4607  & -0.3092  & -0.0631  & -0.6359  & 0.4432  & -0.2956 \tabularnewline
8  & 4  & 0.4709  & 0.6359  & -0.2463  & -0.3092  & -0.4544  & 0.1051 \tabularnewline
8  & 6  & 0.4607  & 0.3092  & -0.0631  & 0.6359  & 0.4432  & -0.2956 \tabularnewline
8  & 8  & 0.2869  & 0.0000  & 0.3343  & 0.0000  & 0.3110  & 0.8421 \tabularnewline
\hline 
\multicolumn{2}{l}{$J=12$} &  &  &  &  &  & \tabularnewline
 &  & 5.0303  & 5.8274  & 6.1835  & 6.7289  & 6.8648  & 9.0079 \tabularnewline
$J_{p}$  & $J_{n}$  &  &  & T=1 & \multicolumn{1}{r}{} & \multicolumn{1}{r}{$T=1$} & \multicolumn{1}{r}{$T=2$}\tabularnewline
4  & 8  & 0.4364  & 0.3052  & -0.3894  & -0.3592  & -0.5903  & 0.2957 \tabularnewline
6  & 6  & 0.7797  & -0.4079  & 0.0000  & 0.2927  & 0.0000  & -0.3742 \tabularnewline
6  & 8  & 0.0344  & 0.5602  & -0.5903  & 0.2078  & 0.3894  & -0.3766 \tabularnewline
8  & 4  & 0.4364  & 0.3052  & 0.3894  & -0.3592  & 0.5903  & 0.2957 \tabularnewline
8  & 6  & 0.0344  & 0.5602  & 0.5903  & 0.2078  & -0.3894  & -0.3766 \tabularnewline
8  & 8  & 0.0940  & 0.1402  & 0.0000  & 0.7550  & 0.0000  & 0.6337 \tabularnewline
\hline 
\multicolumn{2}{l}{$J=13$} &  &  &  &  &  & \tabularnewline
 &  & 5.8951  & 6.1898  & 7.5023  &  &  & \tabularnewline
$J_{p}$  & $J_{n}$  &  & \multicolumn{1}{r}{$T=1$} & \multicolumn{1}{r}{$T=1$} &  &  & \tabularnewline
6  & 8  & 0.7071  & 0.6097  & -0.3581  &  &  & \tabularnewline
8  & 6  & -0.7071  & 0.6097  & -0.3581  &  &  & \tabularnewline
8  & 8  & 0.0000  & 0.5065  & 0.8623  &  &  & \tabularnewline
\hline 
\multicolumn{2}{l}{$J=14$} &  &  &  &  &  & \tabularnewline
 &  & 5.1098  & 6.4980  & 6.7036  &  &  & \tabularnewline
$J_{p}$  & $J_{n}$  &  &  & \multicolumn{1}{r}{$T=1$} &  &  & \tabularnewline
6  & 8  & 0.6943  & -0.1339  & -0.7071  &  &  & \tabularnewline
8  & 6  & 0.6943  & -0.1339  & 0.7071  &  &  & \tabularnewline
8  & 8  & 0.1894  & 0.9819  & 0.0000  &  &  & \tabularnewline
\hline 
\multicolumn{2}{l}{$J=15$} &  &  &  &  &  & \tabularnewline
 &  & 6.2789  &  &  &  &  & \tabularnewline
$J_{p}$  & $J_{n}$  & \multicolumn{1}{r}{$T=1$} &  &  &  &  & \tabularnewline
8  & 8  & 1.0000  &  &  &  &  & \tabularnewline
\hline 
\multicolumn{2}{l}{$J=16$} &  &  &  &  &  & \tabularnewline
 &  & 4.9371  &  &  &  &  & \tabularnewline
$J_{p}$  & $J_{n}$  &  &  &  &  &  & \tabularnewline
8  & 8  & 1.0000  &  &  &  &  & \tabularnewline
\end{tabular}\end{ruledtabular} 
\end{table*}

\begin{table*}[htb]
 \caption{\label{tab:96cd-intd0} Wave functions and energies (in MeV, at the
top) of selected states of $^{96}$Cd calculated with the INTd interaction
(see text) with $T=0$ matrix elements set to zero.}

\begin{ruledtabular} %
\begin{tabular}{cccccccc}
\multicolumn{2}{l}{$J=11$} &  &  &  &  &  & \tabularnewline
 &  & 5.0829  & 5.3798  & 6.8295  & 7.4699  & 7.5178  & 7.8842 \tabularnewline
$J_{p}$  & $J_{n}$  &  &  & \multicolumn{1}{r}{$T=1$} & \multicolumn{1}{r}{$T=1$} & \multicolumn{1}{r}{$T=1$} & \multicolumn{1}{r}{$T=1$}\tabularnewline
4  & 8  & 0.7071  & 0.0000  & 0.2933  & -0.5491  & 0.3351  & -0.0121 \tabularnewline
6  & 6  & 0.0000  & 0.0000  & 0.2913  & 0.5605  & 0.6482  & -0.4253 \tabularnewline
6  & 8  & 0.0000  & 0.7071  & 0.5350  & 0.0396  & -0.4111  & -0.2079 \tabularnewline
8  & 4  & -0.7071  & 0.0000  & 0.2933  & -0.5491  & 0.3351  & -0.0121 \tabularnewline
8  & 6  & 0.0000  & -0.7071  & 0.5350  & 0.0396  & -0.4111  & -0.2079 \tabularnewline
8  & 8  & 0.0000  & 0.0000  & 0.4130  & 0.2822  & 0.1319  & 0.8558 \tabularnewline
\hline 
\multicolumn{2}{l}{$J=12$} &  &  &  &  &  & \tabularnewline
 &  & 5.1165  & 5.2336  & 5.4865  & 7.5293  & 7.5959  & 12.4531 \tabularnewline
$J_{p}$  & $J_{n}$  &  &  &  & \multicolumn{1}{r}{$T=1$} & \multicolumn{1}{r}{$T=1$} & \multicolumn{1}{r}{$T=2$}\tabularnewline
4  & 8  & 0.5699  & 0.2803  & -0.0961  & -0.4783  & 0.5208  & 0.2957 \tabularnewline
6  & 6  & 0.5712  & -0.7151  & 0.1498  & 0.0000  & 0.0000  & -0.3742 \tabularnewline
6  & 8  & 0.0925  & 0.3679  & 0.4629  & -0.5208  & -0.4783  & -0.3766 \tabularnewline
8  & 4  & 0.5699  & 0.2803  & -0.0961  & 0.4783  & -0.5208  & 0.2957 \tabularnewline
8  & 6  & 0.0925  & 0.3679  & 0.4629  & 0.5208  & 0.4783  & -0.3766 \tabularnewline
8  & 8  & -0.0846  & -0.2465  & 0.7284  & 0.0000  & 0.0000  & 0.6337 \tabularnewline
\hline 
\multicolumn{2}{l}{$J=13$} &  &  &  &  &  & \tabularnewline
 &  & 5.3798  & 7.6143  & 7.8873  &  &  & \tabularnewline
$J_{p}$  & $J_{n}$  &  & \multicolumn{1}{r}{$T=1$} & \multicolumn{1}{r}{$T=1$} &  &  & \tabularnewline
6  & 8  & 0.7071  & 0.5265  & -0.4721  &  &  & \tabularnewline
8  & 6  & -0.7071  & 0.5265  & -0.4721  &  &  & \tabularnewline
8  & 8  & 0.0000  & 0.6676  & 0.7445  &  &  & \tabularnewline
\hline 
\multicolumn{2}{l}{$J=14$} &  &  &  &  &  & \tabularnewline
 &  & 5.3798  & 5.6007  & 7.8515  &  &  & \tabularnewline
$J_{p}$  & $J_{n}$  &  &  & \multicolumn{1}{r}{$T=1$} &  &  & \tabularnewline
6  & 8  & 0.7071  & 0.0000  & -0.7071  &  &  & \tabularnewline
8  & 6  & 0.7071  & 0.0000  & 0.7071  &  &  & \tabularnewline
8  & 8  & 0.0000  & 1.0000  & 0.0000  &  &  & \tabularnewline
\hline 
\multicolumn{2}{l}{$J=15$} &  &  &  &  &  & \tabularnewline
 &  & 7.9251  &  &  &  &  & \tabularnewline
$J_{p}$  & $J_{n}$  & \multicolumn{1}{r}{$T=1$} &  &  &  &  & \tabularnewline
8  & 8  & 1.0000  &  &  &  &  & \tabularnewline
\hline 
\multicolumn{2}{l}{$J=16$} &  &  &  &  &  & \tabularnewline
 &  & 5.6007  &  &  &  &  & \tabularnewline
$J_{p}$  & $J_{n}$  &  &  &  &  &  & \tabularnewline
8  & 8  & 1.0000  &  &  &  &  & \tabularnewline
\end{tabular}\end{ruledtabular} 
\end{table*}

Let us first focus on the $J=11^{+}$ and $J=12^{+}$ states. Relative
to Table~\ref{tab:96cd-intd}, we see certain simplicities for the
$J=11^{+}$ states in Table~\ref{tab:96cd-intd0} (where the $T=0$
two-body interaction matrix elements are set to a constant). For the
lowest state, the only non-zero components are $(J_{p},J_{n})=(4,8)$
and $(8,4)$; for the second state, they are $(6,8)$ and $(8,6)$.
This confirms what we said above: $(J_{p},J_{n})$ are good dual quantum
numbers. Nothing special happens to $J=11^{+}$, $T=1$ states.

We show results for $J=12^{+}$ as a counterpoint. We see that nothing
special happens as we go from Table~\ref{tab:96cd-intd} to Table~\ref{tab:96cd-intd0}--no
PDS. The reason for this is, as discussed above, that four identical
$g_{9/2}$ nucleons can have $J=12^{+}$, but, because of the Pauli
Principle, they cannot couple to $J=11^{+}$.

The other states with $J=13,14,15,16$ cannot occur for four identical
nucleons and are therefore subject to the PDS. Note certain degeneracies,
e.g. $J=11,13$, and 14 states, all with $(J_{p},J_{n})=(6,8)$ and
$(8,6)$, have the same energy $E=5.3798$~MeV. The proof of all
these properties are contained in Refs.~\cite{zr11,rz01}.

The $J=16^{+}$, which was experimentally discovered by Nara Singh
\textit{et al}.~\cite{setal11} is correctly predicted to be isomeric
in Table~\ref{tab:96cd-intd}. It lies below the lowest $15^{+}$
or $14^{+}$ states. In Table~\ref{tab:96cd-intd0}, however, the
$J=16^{+}$ state lies above the lowest $J=14^{+}$ state and is degenerate
with the second $J=14^{+}$ state ($E=5.6007$~MeV). Clearly, fluctuations
in the $T=0$ matrix elements are responsible for making the $J=16^{+}$
isomeric.


\begin{thebibliography}{References}
\bibitem{setal11} B.S. Nara Singh \textit{et al.}, Phys. Rev. Lett.
\textbf{107}, 172502 (2011).

\bibitem{zr11} L.Zamick and S.J.Q. Robinson, Phys. Rev. C \textbf{84},
044325 (2011).

\bibitem{rz01} S.J.Q. Robinson and L. Zamick, Phys. Rev. C \textbf{63},
064316 (2001).

\bibitem{ez06} A. Escuderos and L. Zamick, Phys. Rev. C \textbf{73},
044302 (2006).

\bibitem{z07} L. Zamick, Phys. Rev. C \textbf{75}, 064305 (2007).

\bibitem{ezb05} A. Escuderos, L. Zamick, and B.F. Bayman, \textit{Wave
functions in the $f_{7/2}$ shell, for educational purposes and ideas},
\url{http://arxiv.org/abs/nucl-th/0506050} (2005).

\bibitem{ih08} P. Van Isacker and S. Heinze, Phys. Rev. Lett. \textbf{100},
052501 (2008).

\bibitem{zi08} L. Zamick and P. Van Isacker, Phys. Rev. C \textbf{78},
044327 (2008).

\bibitem{q11} Chong Qi, Phys. Rev. C \textbf{83}, 014307 (2011).

\bibitem{qxl12} Chong Qi, Z.X. Xu, and R.J. Liotta, Nucl. Phys. A
\textbf{884--885}, 21--35 (2012).

\bibitem{t10} I. Talmi, Nucl. Phys. A \textbf{846}, 31 (2010).

\bibitem{cetal11} B. Cederwall \textit{et al.}, Nature (London) \textbf{469},
68 (2011).

\bibitem{zi11} S. Zerguine and P. Van Isacker, Phys. Rev. C \textbf{83},
064314 (2011).

\bibitem{qbbcjlw11} C. Qi, J. Blomqvist, T. Bäck, B. Cederwall, A.
Johnson, R.J. Liotta, and R. Wyss, Phys. Rev. C \textbf{84}, 021301(R)
(2011).

\bibitem{xqblw12} Z.X. Xu, C. Qi, J. Blomqvist, R.J. Liotta, and
R. Wyss, Nucl. Phys. A \textbf{877}, 51 (2012).

\bibitem{ze12} L. Zamick and A. Escuderos, Nucl. Phys. A \textbf{889},
8 (2012).

\bibitem{ccgi12} L. Coraggio, A. Covello, A. Gargano, and N. Itaco,
Phys. Rev. C \textbf{85}, 034335 (2012).

\bibitem{zr02} L. Zamick and S.J.Q. Robinson, Phys. Atom. Nucl. \textbf{65},
740 (2002).

\bibitem{bmz63} B.F. Bayman, J.D. McCullen, and L. Zamick, Phys.
Rev. Lett. \textbf{11}, 215 (1963).

\bibitem{gf63} J.N. Ginocchio and J.B. French, Phys. Lett. \textbf{7},
137 (1963). \end{thebibliography}
\end{document}